\documentclass[10pt]{article}
\usepackage{amssymb}
\newtheorem{theorem}{Theorem}

\newtheorem{proposition}{Proposition}

\begin{document}
\title{Unique determination of potentials and semilinear terms of
semilinear elliptic equations from partial Cauchy data}

\author{
O.~Yu.~Imanuvilov\thanks{ Department
of Mathematics, Colorado State University, 101 Weber Building, Fort
Collins, CO 80523-1874, USA, E-mail: oleg@math.colostate.edu}\, and
M.~Yamamoto\thanks{{\bf contacting author}:
Department of Mathematical Sciences, The University
of Tokyo, Komaba Meguro Tokyo 153-8914 Japan,
E-mail:myama@ms.u-tokyo.ac.jp,
tel: +81-3-5465-7011, fax: +81-3-5465-7011}
}

\date{}
\maketitle

\begin{abstract} For a semilinear elliptic equation, we prove
uniqueness results in determining potentials and semilinear terms
from partial Cauchy data on an arbitrary subboundary.
\end{abstract}

\section{Introduction}\label{sec1}

Let $\Omega$ be a bounded domain in $\Bbb R^2$ with smooth boundary,
$\widetilde \Gamma$ be a relatively open subset on $\partial\Omega$ and
$\Gamma_0 = \partial\Omega\setminus {\overline{\widetilde \Gamma}}.$

Consider the following boundary value problem:
\begin{equation}\label{o-1}
P(x,D)u := \Delta u + q(x)u - f(x,u) =0 \quad
\mbox{in}\,\,\Omega,\quad  u\vert_{\Gamma_0}=0,
\end{equation}
Henceforth we set $L(x,D)u = \Delta u+qu .$

Consider the following partial Cauchy data:
$$
\mathcal C_{q,f} = \left\{\left(u,\frac{\partial u}{\partial
\nu}\right)\biggl\vert_{\widetilde \Gamma}\thinspace ; \thinspace
P(x,D)u=0\quad\mbox{in}\,\,\Omega,\quad u\vert_{\Gamma_0}=0,\,\,
u\in H^1(\Omega)\right\}.
$$
Here $\nu$ is the unit outward normal vector to $\partial\Omega$.

The paper is concerned with the following inverse problem: {\it
Using the partial Cauchy data $\mathcal C_{q}$, determine the
coefficient $q$. }
Assume that
\begin{equation}\label{00}
f, \frac{\partial f}{\partial y}, \frac{\partial^2 f}{\partial
y^2}\in C^0(\overline\Omega\times {\Bbb R^1}), \quad
f(x,0)\equiv\frac{\partial f(x,0)}{\partial y}\equiv 0,
\end{equation}
and for some positive constants $p>1, C_1, C_2$, the following holds
true:
\begin{equation}\label{01}
f(x,y)y\ge C_1 \vert
y\vert^{p+1}-C_2, \quad \forall (x,y)\in\Omega\times {\Bbb R}^1.
\end{equation}
Moreover for some $p_1>0, p_2>0$, $C_3>0$ and $C_4>0$, the following
inequalities holds true:
\begin{equation}\label{02}
\vert \frac{\partial f}{\partial y} (x,y)\vert\le
C_3(1+\vert y\vert^{p_1}),\quad \forall (x,y)\in\Omega\times {\Bbb
R}^1, \quad \vert \frac{\partial^2 f}{\partial y^2} (x,y)\vert\le
C_4(1+\vert y\vert^{p_2}),\quad \forall (x,y)\in\Omega\times {\Bbb
R}^1.
\end{equation}

Our first main result is concerned with the uniqueness in
determining a linear part, that is, a potential $q$.
\begin{theorem}\label{vokal}
Let functions $f_1,f_2$ satisfy (\ref{00}), (\ref{01}), (\ref{02})
and $q_j\in C^{2+\alpha}(\overline \Omega)$, $j=1,2$, with some
$\alpha\in (0,1).$ Suppose that $\mathcal C_{q_1,f_1}=\mathcal C_{q_2,f_2}$.
Then $q_1=q_2$ in $\Omega$.
\end{theorem}

{\bf Remark 1.} {\it Since our assumptions on the potential $q$ and
nonlinear term $f$ in general do not imply the uniqueness of a
solution for the boundary value problem for the elliptic operator
$P(x,D)$, by the equality $\mathcal C_{q_1,f_1}=\mathcal C_{q_2,f_2}$,
we mean the following: for any element $(v_1,v_2)$ from
$\mathcal C_{q_1,f_1}$, there exists a function $w\in H^1(\Omega)$
such that $P_2(x,D)w=\Delta w+q_2w-f_2(x,w)=0,
w\vert_{\Gamma_0}=0, w\vert_{\widetilde \Gamma}=v_1$ and
$\frac{\partial w}{\partial \nu}\vert_{\widetilde \Gamma}=v_2.$ }
\\
\vspace{0.4cm}
\\
{\bf Remark 2.} {\it Theorem
\ref{vokal} is still true if condition (\ref{01}) is
replaced by following: there exists a continuous function $G$
such that a solution to the boundary value problem
$$
P(x,D)u=0\quad \mbox{in}\,\,\Omega, \quad u\vert_{\partial\Omega}=g
$$
satisfies the estimate
$$
\Vert u\Vert_{H^1(\Omega)}\le G(\Vert g\Vert_{H^\frac 12(\partial\Omega)}).
$$
Condition (\ref{01}) is used in deriving the inequality
(\ref{(2.1II)}).
}
\\
\vspace{0.4cm}
\\
For any $F(t)\in C([0,1];C^{2+\alpha}(\overline\Omega))$
with $\alpha\in (0,1)$, we introduce the set
$$
\mathcal O_F=\bigcup_{0\le t\le 1, x\in \Omega}
\{(x,(F(t))(x))\}.
$$
Let
\begin{eqnarray*}
&&\mathcal U_j=\{F\in C([0,1];C^{2+\alpha}(\overline\Omega)); \thinspace
F(0)=0 \quad \mbox{$u(\cdot,t):= F(t)$ satisfies}\\
&& \Delta u(x,t) + q_ju(x,t) - f_j(x, u(x,t)) = 0,
\thinspace x \in \Omega, \quad u(\cdot,t)\vert_{\Gamma_0}=0\}, \quad
j=1,2.
\end{eqnarray*}
\\
The second main result asserts the uniqueness for semilinear
terms $f_k$, $k=1,2$ in some range provided that the potential $q$
is known:
\\
\vspace{0.3cm}
\begin{theorem}\label{vokall}
Let $q_1=q_2=q \in C^{2+\alpha}(\overline{\Omega})$ be arbitrarily fixed.
Let functions $f_1,f_2\in C^{3+\alpha}(\overline\Omega\times \Bbb R^1)$
for some $\alpha\in (0,1)$, satisfy
(\ref{01}), (\ref{02}) and $f_1(\cdot,0) = f_2(\cdot,0)=0$.
Suppose that $\mathcal C_{q,f_1}=\mathcal C_{q,f_2}.$
Then
$$
f_1-f_2 = 0 \quad \mbox{in $\bigcup_{j\in\{ 1,2\}}
\bigcup_{F\in \mathcal U_j} \mathcal O_{F}$}.
$$
\end{theorem}

Combining Theorems 1 and 2, under stronger conditions on $f_k$, $k=1,2$,
we can prove the uniqueness in determining both $q$ and $f(x,u)$.
\\
{\bf Corollary}
{\it Let $q_1, q_2 \in C^{2+\alpha}(\overline{\Omega})$ and let
functions $f_1,f_2\in C^{3+\alpha}(\overline\Omega\times \Bbb R^1)$
with some $\alpha\in (0,1)$, satisfy (\ref{00}), (\ref{01})
and (\ref{02}).
Suppose that $\mathcal C_{q_1,f_1}=\mathcal C_{q_2,f_2}.$
Then $q_1 = q_2$ in $\Omega$ and
$$
f_1-f_2 = 0 \quad \mbox{in $\bigcup_{j\in\{ 1,2\}}
\bigcup_{F\in \mathcal U_j} \mathcal O_{F}$}.
$$
}
\\
\vspace{0.3cm}
{\bf Remark 3.}
{\it Under the condition of Theorem \ref{vokal}, we can not completely
recover the nonlinear term.
Indeed, if $\rho\in C^2(\overline\Omega)$, $\rho\vert_{\partial\Omega}=0$,
$\frac{\partial\rho}{\partial\nu}<0$ on $\partial\Omega$ and
$\rho>0$ in $\Omega,$ under assumptions (\ref{00}) and (\ref{01}),
we have the following a priori estimate proved in \cite{FI}:
$$
\int_\Omega \rho^{\kappa}(\vert \nabla u\vert^2+\vert u\vert^{p+1})dx
\le C
$$
for $u \in H^1(\Omega)$ satisfying $P(x,D)u=0$ in $\Omega$.
Here a constant $C$ is independent of $u$ and $\kappa$ depends on $p.$
Such a estimate immediately implies that for any
$\Omega_1\subset\subset \Omega$, there exists a constant $C(\Omega_1)>0$
such that
$$
\Vert u\Vert_{C^0(\overline{\Omega_1})}\le C(\Omega_1).
$$
This estimate and (\ref{01}) imply that for any $x\in \Omega_1$  a nonlinear term $f(x,y)$
can not be recovered for all sufficiently large $y$.
}
\vspace{0.4cm}

Restricted to the linear elliptic equation in two dimensions, there are
quite rich references and here we give a very partial list.
In the case $\widetilde \Gamma =\partial\Omega$ of the full Cauchy
data, the uniqueness in determining a potential $q$ in the two
dimensional case was proved for the conductivity equation by Nachman
in \cite{N} within $C^4$ conductivities, and later in \cite {AP}
within $L^\infty$ conductivities.  For a convection equation, see
\cite{ChengYama}. The case of the Schr\"odinger equation was solved
by Bukhegim \cite {Bu}. In the case of the partial Cauchy data on
arbitrary subboundary, the uniqueness was obtained in \cite{IUY} for
potential $q \in C^{2+\alpha}(\overline\Omega)$, and in \cite {IY},
the regularity assumption was improved to
$C^\alpha(\overline\Omega)$ in the case of the full Cauchy data and
up to $W^1_p(\Omega)$ with $p>2$ in the case of partial Cauchy data
on arbitrary subboundary. The case of general second-order elliptic
equation was studied in the papers \cite{IUY1} and \cite{IUY2}. The
results of \cite{IUY} were extended to a Riemannian surface in
\cite{GT} and system of linear equations in \cite{IY1}. The case
where voltages are applied and currents are measured on disjoint
subboundaries was discussed and the uniqueness is proved in
\cite{IUY3}.   Conditional stability estimates in determining a
potential are obtained in \cite{Nov}.  As for the cases of
the dimensions $\ge 3$, we refer to \cite{KSU} and the
references therein.

Our main results establish the uniqueness in determining
semilinear terms by partial Cauchy data
on arbitrary subboundary, and to our best knowledge, there are
no publications in this case.
On the other hand, we can refer to several works on nonlinear elliptic
equations by not arbitrary subboundary as follows.
The uniqueness results for recovery of the nonlinear term in the
semilinear elliptic equation were  first obtained for the full
Cauchy data in three dimensional case  by Isakov and Sylvester
in \cite{IN} and in two dimensional case by Isakov and Nachman in \cite{INa}.
It should be mentioned that the proof of the analog of Theorem 1 in those
papers requires the  uniqueness of solution  for the Dirichlet boundary
problem for the operator $P(x,D)$.
Later this result was expanded to the case of
a system of semilinear elliptic equations by Isakov in \cite{IS2}.
Also see Kang and Nakamura \cite{KN} for determination of coefficients
of the linear and the quadratic nonlinear terms in the principal
part of a quasilinear elliptic equation.
In a special case where a nonlinear term is
independent of $x$, the uniqueness was proved in determining such a
nonlinear term from partial Cauchy data
\cite{IS1}.  Moreover we note that in \cite{IS2} and
\cite{IN}, the monotonity of $f(x,u)$ with respect to $u$ is assumed.
In general, if a nonlinear term depends on  $x$, $u$ and
the gradient of $u$, then it is impossible to prove the uniqueness
even for the linear case.
This can be seen by \cite {IY1} if we consider
the term $-f(x,u,\nabla u) = A(x)\cdot \nabla u
+ q(x)u$ .

The paper is composed of four sections.  In section 2, we prove
Theorem 2 provided that Theorem 1 is proved.
Sections 3 and 4 are devoted to the proof of Theorem 1.

\section{Proof of Theorem 2}\label{sec1}

Henceforth let $\partial_x^\beta = \partial_{x_1}^{\beta_1}
\partial_{x_2}^{\beta_2}$, $\beta=(\beta_1,\beta_2) \in (\Bbb N \cup\{0\})^2$
and $\vert \beta\vert=\beta_1+\beta_2$.
We set $P_k(x,D)u = \Delta u + q(x)u - f_k(x,u)$, $k=1,2$, and
$u_{1,t}(x)=u(x,t)$.
Let $u_{2,t}\in H^1(\Omega)$, $t\in [0,1]$ satisfy
$$
P_2(x,D)u_{2,t}=0\quad\mbox{in}\,\,\Omega, \quad
u_{2,t} = u_{1,t} \quad \mbox{on $\partial\Omega$},
\quad \forall t \in [0,1].
$$
Then $\mathcal{C}_{q,f_1} = \mathcal{C}_{q,f_2}$ yields
$$
\left( \frac{\partial u_{1,t}}{\partial\nu}
- \frac{\partial u_{2,t}}{\partial\nu}\right)
\vert_{\widetilde\Gamma}=0, \quad \forall t\in [0,1].
$$
By (\ref{02}) and the Sobolev embedding theorem,
$f_2(\cdot,u_{2,t}(\cdot))\in L^{\kappa}(\Omega)$ for any $\kappa>1.$
The standard solvability theory for the Dirichlet boundary value problem
for the Laplace operator in Sobolev spaces implies $u_{2,t}\in H^2(\Omega).$
Hence $f_2(\cdot,u_{2,t}(\cdot))\in C^{\widetilde\alpha}(\Omega)$ for any
$\widetilde \alpha\in (0,1).$
Then, since $u_{2,t}\in C^{2+\alpha}(\partial\Omega)$,
the solvability theory for the Dirichlet boundary value problem for
the Laplace operator in H\"older spaces implies
$u_{2,t}\in C^{2+\alpha}(\overline\Omega).$
By the assumption, there exists a constant $K>0$ such that
\begin{equation}\label{-10}
\sup_{t\in [0,1]}\Vert u_{1,t}\Vert_{C^0(\overline\Omega)}\le K.
\end{equation}

We claim that
\begin{equation}\label{eb1}
u_{1,t}\equiv u_{2,t}, \quad \forall t\in [0,1].
\end{equation}
Our proof is by contradiction. Suppose that for some $t_0\in (0,1]$,
this equality fails. Let $t_*$ be the infimum over such $t_0.$

Setting $u_t=u_{2,t}-u_{1,t}$, we have
\begin{equation}
\Delta u_t - q(t,x)u_t=f_1(x,u_{1,t})-f_2(x,u_{1,t})\quad
\mbox{in}\,\,\Omega,\quad u_{t}\vert_{\partial \Omega}=0,
\quad\frac{\partial u_t}{\partial\nu}\vert_{\widetilde \Gamma}=0,
\end{equation}
where $q(t,x) = -q(x) + \int_0^1\frac{\partial f_2}{\partial y}(x,
(1-s)u_{2,t}(x)+su_{1,t}(x))ds$.   To this equation, applying
a Carleman estimate with boundary term (see e.g., \cite{IP}),
we can choose some function $\phi \in C^2(\overline{\Omega})$ such that
$$
\Vert e^{\tau\phi} u_t\Vert_{H^{2,\tau}(\Omega)}\le C\tau^{\frac
32}\Vert e^{\tau\phi}
(f_1(\cdot,u_1)-f_2(\cdot,u_1))\Vert_{L^2(\Omega)},\quad \forall \tau\ge
\tau_0.
$$
Here $\Vert v\Vert_{H^{2,\tau}(\Omega)} = \left(\sum_{\vert \beta\vert \le 2}
\tau^{4-2\vert\beta\vert} \Vert v\Vert_{H^{\vert\beta\vert}(\Omega)}^2
\right)^{\frac{1}{2}}$.
That is,  fixing a large $\tau > 0$ arbitrarily,
\begin{equation}\label{k1}
\Vert u_t\Vert_{H^2(\Omega)}\le C
\Vert f_1(\cdot,u_1)-f_2(\cdot,u_1)\Vert_{L^2(\Omega)},\quad \forall
t \in [0,1],
\end{equation}
where a constant $C > 0$ depends on fixed $\tau$.

Consider the boundary value problem
\begin{eqnarray*}
&&\Delta v_{k,t} + q(x)v_{k,t}
- \frac{\partial f_k}{\partial y}(x,u_{k,t})v_{k,t}
-\tilde f_k(x,v_{k,t})\\
&=& \Delta v_{k,t} + q(x)v_{k,t} - f_k(x,v_{k,t}+u_{k,t})
+ f_k(x,u_{k,t}) =0\quad\mbox{in}\,\,\Omega,\quad
v_{k,t}\vert_{\Gamma_0}= 0,
\end{eqnarray*}
where $\widetilde f_k(x,w)=f_k(x,w+u_{k,t})-f_k(x,u_{k,t})
-\frac{\partial f_k}{\partial y}(x,u_{k,t})w.$
Obviously the functions $\tilde f_k$ satisfy (\ref{00}), (\ref{01})
and (\ref{02}).
Moreover
$$
\mathcal C_{q-\frac{\partial f_1}{\partial y}(x,u_{1,t}),\tilde f_1}
=\mathcal C_{q-\frac{\partial f_2}{\partial y}(x,u_{2,t}),\tilde f_2}.
$$

Indeed let $(w_1,w_2)\in
\mathcal C_{q- \frac{\partial f_1}{\partial y}(x,u_{1,t}),\tilde f_1}$.
Let $w \in H^1(\Omega)$ be the solution to the boundary value problem
$$
\Delta w + qw - \frac{\partial f_1}{\partial y}(x,u_{1,t})w
-\tilde f_1(x,w)=0\quad\mbox{in}\,\,\Omega,\quad
w\vert_{\Gamma_0}= 0,\quad w\vert_{\tilde \Gamma}=w_1.
$$
such that $\frac{\partial w}{\partial\nu}\vert_{\tilde \Gamma}=w_2$.

On the other hand, the function $w+u_{1,t}$ solves the boundary value problem
$$
\Delta(w+u_{1,t}) + q(w+u_{1,t}) - f_1(x, w+u_{1,t})=0\quad\mbox{in}
\,\,\Omega,\quad (w+u_{1,t})\vert_{\Gamma_0}= 0.
$$
Let $\widetilde u$ satisfy
$$
\Delta \widetilde{u} + q\widetilde{u} - f_2(x, \widetilde u) =0
\quad\mbox{in}\,\,\Omega,\quad \widetilde u\vert_{\Gamma_0}= 0
$$
and
$$
\widetilde{u} = w + u_{1,t} \quad \mbox{on $\widetilde{\Gamma}$}.
$$
Then, by assumption $\mathcal{C}_{q,f_1} = \mathcal{C}_{q,f_2}$, we have
$$
\frac{\partial\widetilde{u}}{\partial\nu}
= \frac{\partial(w+u_{1,t})}{\partial\nu} \quad \mbox{on
$\widetilde{\Gamma}$}.
$$
Setting $\widetilde w = \widetilde u - u_{2,t}$, we obtain
$$
\Delta\tilde  w + q\widetilde w
- \frac{\partial f_2}{\partial y}(x,u_{2,t})\tilde w
-\tilde f_2(x,\tilde w)=0\quad\mbox{in}\,\,\Omega,\quad
\widetilde w\vert_{\Gamma_0}= 0.
$$
Then on $\tilde \Gamma$ we have
$$
\widetilde{w} - w = (\widetilde{u} - u_{2,t})
- (\widetilde u - u_{1,t}) = u_{1,t} - u_{2,t} = 0
$$
and
\begin{eqnarray*}
&& \frac{\partial \widetilde w}{\partial\nu}
- \frac{\partial w}{\partial\nu}
= \frac{\partial \widetilde u}{\partial\nu}
- \frac{\partial u_{2,t}}{\partial\nu}
- \frac{\partial w}{\partial\nu}\\
&=& \frac{\partial w}{\partial\nu} + \frac{\partial u_{1,t}}{\partial\nu}
- \frac{\partial u_{2,t}}{\partial\nu}
- \frac{\partial w}{\partial\nu} = 0.
\end{eqnarray*}
Therefore $\widetilde w = w_1$ and $\frac{\partial \widetilde w}
{\partial\nu} = w_2$ on $\widetilde{\Gamma}$.  Hence
$(w_1,w_2) \in \mathcal C_{q-\frac{\partial f_2}{\partial y}(x,u_{2,t}),
\tilde f_2}$.  Since the reverse inclusion can be proved similarly, we
have proved
$$
\mathcal C_{q-\frac{\partial f_1}{\partial y}(x,u_{1,t}),\tilde f_1}
=\mathcal C_{q-\frac{\partial f_2}{\partial y}(x,u_{2,t}),\tilde f_2}.
$$

\vspace{0.4cm}

Therefore we can apply Theorem 1 to this equation.  Hence we have the
uniqueness for the potential, that is,
\begin{equation} \label{eb}
\frac{\partial f_1}{\partial y}(x,u_{1,t})
= \frac{\partial f_2}{\partial y}(x, u_{2,t}) \quad\mbox{in}\,\,\Omega,
\quad \forall t\in [0,1].
\end{equation}
Denote $\gamma(t)=\Vert
u_{1,t}-u_{1,t_*}\Vert_{C^0(\overline\Omega)}+\Vert
u_{2,t}-u_{2,t_*}\Vert_{C^0(\overline\Omega)}.$
Since $u_{1,t_*} =u_{2,t_*}$ in $\Omega$, we have
$f_1(x,u_{1,t_*})=\Delta u_{1,t_*} = \Delta u_{2,t_*}
= f_2(x,u_{1,t_*})$ in $\Omega.$ Therefore
$$
f_1(x, u_{1,t}(x))-f_2(x,u_{1,t}(x))=\int_{{u_{1,t_*}(x)}}^{u_{1,t}(x)}\left (
\frac{\partial f_1}{\partial y}(x,s)-\frac{\partial f_2}{\partial y}(x,s)\right )
ds.
$$
If $s\in (u_{1,t_*}(x),u_{1,t}(x))$, then, by the continuity of
$u_{1,t}(x)$ with respect to $t$ and the intermediate value theorem,
there exists $t_0(s,x)\in [0,t]$ such that $s=u_{1,t_0(s,x)}(x).$ Hence
$$
f_1(x, u_{1,t}(x))-f_2(x,u_{1,t}(x))
= \int_{{u_{1,t_*}(x)}}^{u_{1,t}(x)}\left(
\frac{\partial f_1}{\partial y}(x,u_{1,t_0(s,x)}(x))
- \frac{\partial f_2}{\partial y}(x,u_{1,t_0(s,x)}(x))\right)ds.
$$
Applying (\ref{eb}) and (\ref{-10}), we have
\begin{eqnarray}\label{ooooo}
f_1(x, u_{1,t}(x))-f_2(x,u_{1,t}(x))=\int_{{u_{1,t_*}(x)}}^{u_{1,t}(x)}
\left(\frac{\partial f_2}{\partial y}(x,u_{2,t_0(s,x)}(x))
- \frac{\partial f_2}{\partial y}(x,u_{1,t_0(s,x)}(x))\right)ds
                                                 \nonumber\\
\le \left\Vert \frac{\partial^2 f_2}{\partial
y^2}\right\Vert_{C^0(\overline\Omega\times [-K,K])} \sup_{\widetilde t\in
(0,t)}\vert (u_{1,\widetilde t}-u_{2,\widetilde t})(x)\vert \gamma( t)\nonumber\\
\le \left\Vert \frac{\partial^2 f_2}{\partial
y^2}\right\Vert_{C^0(\overline\Omega\times [-K,K])} \sup_{\widetilde t\in
(t_*,t)}\vert (u_{1,\widetilde t}-u_{2,\widetilde t})(x)\vert \gamma( t).
\end{eqnarray}
In order to obtain the last inequality, we used the fact that
$u_{1,\widetilde t}-u_{2,\widetilde t}\equiv 0$ for all $\widetilde t$ from $[0,t_*].$
Therefore inequality (\ref{ooooo}) implies
\begin{equation}\label{k2}
\sup_{\widetilde t\in (t_*,t)}\Vert f_1(x, u_{1,\widetilde
t})-f_2(x,u_{1,\widetilde t})\Vert_{L^2(\Omega)}\le C\gamma(t)
\sup_{\widetilde t\in (t_*,t)}\Vert u_{1,\widetilde t}-u_{2,\widetilde
t}\Vert_{L^2(\Omega)}.
\end{equation}
From (\ref{k1}) and (\ref{k2}), we obtain
$$
\Vert u_t\Vert_{H^{2}(\Omega)}\le  C\gamma(t) \sup_{\widetilde t\in
(t_*,t)}\Vert u_{1,\widetilde t}-u_{2,\widetilde t}\Vert_{L^2(\Omega)},
\quad \widetilde{t}\in (t_*,t).
$$
This implies that
\begin{equation} \label{popka}
\sup_{\widetilde t\in
(t_*,t)}\Vert u_{\widetilde t}\Vert_{H^{2}(\Omega)}\le  C\gamma(t)
\sup_{\widetilde t\in (t_*,t)}\Vert u_{\widetilde t} \Vert_{L^2(\Omega)}.
\end{equation}
From (\ref{popka}) and the fact that $\gamma(t)$ goes to zero as
$t\rightarrow t_*$, we obtain that there exists $\hat t>t_*$ such that
$u_{1,t}=u_{2,t}$ for all $t$ from $(t_*,\hat t)$. We arrive at the
contradiction.  Equality (\ref{eb1}) is proved and the statement
of the theorem follows from it and (\ref{eb}).  $\blacksquare$

\section{Preliminaries for the proof of Theorem 1}\label{sec1}

Henceforth we use the following notations.
\\

\noindent {\bf Notations.} $i=\sqrt{-1}$, $x_1, x_2, \xi_1, \xi_2
\in {\Bbb R}^1$, $z=x_1+ix_2$, $\zeta=\xi_1+i\xi_2$, $\overline{z}$
denotes the complex conjugate of $z \in \Bbb C$. We identify $x =
(x_1,x_2) \in {\Bbb R}^2$ with $z = x_1 +ix_2 \in {\Bbb C},$
$\partial_z = \frac 12(\partial_{x_1}-i\partial_{x_2})$,
$\partial_{\overline z}= \frac12(\partial_{x_1}+i\partial_{x_2}),$
$D=(\frac{1}{i}\frac{\partial}{\partial
x_1},\frac{1}{i}\frac{\partial}{\partial x_2}).$ The tangential
derivative on the boundary is given by
$\partial_{\vec\tau}=\nu_2\frac{\partial}{\partial x_1}
-\nu_1\frac{\partial}{\partial x_2},$ where $\nu=(\nu_1, \nu_2)$ is
the unit outer normal to $\partial\Omega.$
 We set $(u,v)_{L^2(\Omega)} =
\int_{\Omega} u {v} dx$ for functions $u, v$, while by $(a,b)$ we
denote the scalar product in $\Bbb R^2$ if there is no fear of
confusion. For $f:{\Bbb R}^2\rightarrow {\Bbb R}^1$, the symbol
$f''$ denotes the Hessian matrix  with entries $\frac{\partial^2
f}{\partial x_k\partial x_j},$ $\mathcal L(X,Y)$ denotes the Banach
space of all bounded linear operators from a Banach space $X$ to
another Banach space $Y$. Let $\Vert \cdot\Vert_X$ be the norm in
a Banach space $X$.   We set $ \Vert
u\Vert_{H^{k,\tau}(\Omega)} = (\Vert u\Vert_{H^k(\Omega)}^2 + \vert
\tau\vert^{2k}\Vert u\Vert^2_{L^2(\Omega)})^{\frac{1}{2}}$. By
$o_X(\frac{1}{\tau^\kappa})$ we denote a function $f(\tau,\cdot)$
such that $ \Vert f(\tau,\cdot)\Vert_X=o(\frac{1}{\tau^\kappa})\quad
\mbox{as} \,\,\vert \tau\vert\rightarrow +\infty. $\vspace{0.4cm}

Let $\Omega_*$ be a bounded domain in $\Bbb R^2$ such that $\Omega\subset\Omega_*, \Gamma_0\subset \partial\Omega_*$ and $\widetilde \Gamma\cap \partial\Omega_*=\emptyset.$

For some $\alpha\in (0,1)$, we consider a function
$\Phi(z)=\varphi(x_1,x_2)+i\psi(x_1,x_2) \in
C^{6+\alpha}(\overline{\Omega_*})$ with real-valued $\varphi$ and
$\psi$ such that
\begin{equation}\label{zzz}
\partial_z\Phi(z) = 0 \quad \mbox{in}
\,\,\Omega_*, \quad\mbox{Im}\,\Phi\vert_{\Gamma_0}=0.
\end{equation}
 Denote by $\mathcal H$ the set
of all the critical points of the function $\Phi$:
$$
\mathcal H = \{z\in\overline\Omega_*; \thinspace
\frac{\partial\Phi}{\partial z} (z)=0\}.
$$
Assume that $\Phi$ has no critical points on
$\overline{\widetilde\Gamma}$, and that all critical points  are
nondegenerate:
\begin{equation}\label{mika}
\mathcal H\cap \partial\Omega\subset\Gamma_0,\quad
\partial^2_{z}\Phi (z)\ne 0, \quad \forall z\in
\mathcal H.
\end{equation}
Then $\Phi$  has only a finite number of critical points and we can
set:
\begin{equation}\label{mona}
{\mathcal H} \setminus \Gamma_0= \{ \widetilde{x}_1, ...,
\widetilde{x}_{\ell} \},\quad \mathcal H \cap \Gamma_0=\{
\widetilde{x}_{\ell+1}, ..., \widetilde{x}_{\ell+\ell'} \}.
\end{equation}

Let $\partial \Omega=\cup_{j=1}^{\mathcal N}\gamma_j,$ where
$\gamma_j$ is a closed contour. The following proposition was proved
in \cite{IUY}.

\begin{proposition}\label{Proposition -1}
Let $\widetilde x$ be an arbitrary point in $\Omega.$ There exists a
sequence of functions $\{\Phi_\epsilon\}_{\epsilon\in(0,1)}$
satisfying (\ref{zzz}), (\ref{mika}) and there exists a sequence
$\{\widetilde x_\epsilon\}, \epsilon\in (0,1)$ such that
$$
\widetilde x_\epsilon \in \mathcal H_\epsilon
= \{z\in\overline\Omega; \thinspace
\frac{\partial \Phi_\epsilon}{\partial
z}(z)=0 \},\quad \widetilde x_\epsilon\rightarrow \widetilde
x\,\,\mbox{ as}\,\, \epsilon\rightarrow +0.
$$
Moreover for any $j$ from $\{1,\dots,\mathcal N\}$, we have
$$
\mathcal H_\epsilon\cap\gamma_j=\emptyset\quad\mbox{if}\,\,
\gamma_j\cap \widetilde \Gamma\ne\emptyset,
$$
$$
\mathcal H_\epsilon\cap\gamma_j\subset \Gamma_0 \quad\mbox{if}\,\,
\gamma_j\cap \widetilde \Gamma = \emptyset
$$
and
$$
\mbox{Im}\,\Phi_\epsilon(\widetilde x_\epsilon)\notin \{\mbox{Im}\,
\Phi_\epsilon(x); \thinspace x\in \mathcal H_\epsilon\setminus
\{\widetilde{x_\epsilon}\}\} \,\,\mbox{and}
\,\,\mbox{Im}\,\Phi_\epsilon(\widetilde x_\epsilon) \ne 0.
$$
\end{proposition}

Later we use the following proposition (see \cite{IUY}) :

\begin{proposition}\label{gandonnal} Let $\Phi$ satisfy (\ref{zzz})
and (\ref{mika}). For every $g\in L^1(\Omega)$, we have
$$
\int_\Omega ge^{\tau(\Phi-\overline \Phi)}dx\rightarrow \,\,0\quad
\mbox{as}\quad \tau\rightarrow +\infty .
$$
\end{proposition}

Consider the boundary value problem
$$
L(x,{D})u = f \quad \mbox{in} \quad \Omega, \quad u \vert_{\partial
\Omega} = 0.
$$
The following proposition is proved in \cite{IUY1}.
\begin{proposition}\label{Theorem 2.1}
Suppose that $\Phi$ satisfies (\ref{zzz}) and (\ref{mika}), $u\in
H^1_0(\Omega)$ and $\Vert q\Vert_{L^\infty(\Omega)}\le K$. Then
there exist $\tau_0=\tau_0(K,\Phi)$ and $C=C(K,\Phi)$, independent
of $u$ and $\tau$, such that
\begin{eqnarray}\label{suno4} \vert \tau\vert\Vert
ue^{\tau\varphi}\Vert^2_{L^2(\Omega)}+\Vert
ue^{\tau\varphi}\Vert^2_{H^1(\Omega)}+\Vert\frac{\partial
u}{\partial\nu}e^{\tau\varphi}\Vert^2_{L^2(\Gamma_0)}+
\tau^2\Vert\vert\frac{\partial\Phi}{\partial z} \vert
ue^{\tau\varphi}\Vert^2_{L^2(\Omega)}\nonumber \\
\le C(\Vert (L(x,D)
u)e^{\tau\varphi}\Vert^2_{L^2(\Omega)}+\vert\tau\vert
\int_{\widetilde\Gamma}\vert \frac{\partial
u}{\partial\nu}\vert^2e^{2\tau\varphi}d\sigma)\quad \forall
\vert\tau\vert>\tau_0.
\end{eqnarray}

\end{proposition}

Using estimate (\ref{suno4}), we obtain
\begin{proposition} \label{vanka} Let $\Phi$ satisfy (\ref{zzz})
and (\ref{mika}). There exists a constant $\tau_0$ such that for
$\vert \tau\vert\ge \tau_0$ and any $f\in L^2(\Omega)$ and
$g\in H^{\frac 32}(\partial\Omega)$, there exists
a solution to the boundary value problem:
\begin{equation}\label{lola}
L(x,D+i\tau\nabla\varphi)u =f\quad\mbox{in}
\,\,\Omega, \quad u\vert_{\Gamma_0}=g
\end{equation}
such that
\begin{equation}\label{2}
\Vert u\Vert_{H^{2,\tau}(\Omega)} \le C (\vert\tau\vert^\frac 32\Vert
f\Vert_{L^2(\Omega)}+\Vert g\Vert_{H^\frac 32(\Gamma_0)}).
\end{equation}
The constant $C$ in (\ref{2}) is independent of $\tau.$
\end{proposition}

{\bf Proof.}
First we reduce the problem (\ref{lola})  to the case $g=0.$ Let
$r(z)$ be a holomorphic function and $\widetilde r(\overline z) $ be
an antiholomorphic function  such that $(r+ \widetilde r)\vert_{\Gamma_0}=g.$  These functions can be
chosen such that
$$
\Vert r\Vert_{H^2(\Omega)}+\Vert\widetilde r\Vert_{H^2(\Omega)} \le
C_{24}\Vert g\Vert_{H^\frac 32(\Gamma_0)}.
$$

We look for a solution $u$ in the form
$$
u=(e^{\tau\psi}r +e^{-\tau\psi}\widetilde r)+\widetilde u,
$$
where
\begin{equation}\label{loko}
{ L}(x,D+i\tau\nabla\varphi)\widetilde u=\widetilde f \quad \mbox{in}
\,\,\Omega_*,\quad \widetilde u\vert_{\Gamma_0}=0
\end{equation}
and $\widetilde f= f-qre^{i\tau\psi}-q\widetilde re^{-i\tau\psi}$ is
extended by zero on $\Omega_*\setminus\Omega.$
By Proposition 2.1 of \cite{IUY} there exists a solution to
the problem (\ref{loko}) such that
\begin{equation}
\tau^\frac 12\Vert\widetilde  u\Vert_{L^2(\Omega_*)}
\le C\Vert \widetilde f \Vert_{L^2(\Omega)}
\end{equation}
Obviously the restriction of the function $\widetilde u$ on $\Omega$
satisfies the estimate
\begin{equation}
\Vert\widetilde  u\Vert_{H^{2,\tau}(\Omega)}
\le C\vert\tau\vert^\frac 32\Vert \widetilde f \Vert_{L^2(\Omega)}.
\end{equation}

The proof of the proposition is finished.
$\blacksquare$

Let us introduce the operators:
$$
\partial_{\overline z}^{-1}g=-\frac 1\pi\int_\Omega
\frac{g(\xi_1,\xi_2)}{\zeta-z}d\xi_1d\xi_2,\quad
\partial_{ z}^{-1}g=-\frac 1\pi\int_\Omega
\frac{g(\xi_1,\xi_2)}{\overline\zeta-\overline z}d\xi_1d\xi_2
$$
and
$$
\widetilde {\mathcal R}_{\tau}=\frac 12e^{\overline\Phi-\Phi}
\partial_{ z}^{-1} e^{\Phi-\overline\Phi}, \quad
{\mathcal R}_{\tau}=\frac 12e^{\Phi-\overline\Phi}
\partial_{ z}^{-1} e^{\overline\Phi-\Phi}.
$$

Then we have (e.g., p.47, 56, 72 in \cite{VE}):
\begin{proposition}\label{Proposition 3.0}
{\bf A)} Let $m\ge 0$ be an integer number and $\alpha\in (0,1).$
Then $\partial_{\overline z}^{-1},\partial_{ z}^{-1}\in \mathcal
L(C^{m+\alpha}(\overline{\Omega}),C^{m+\alpha+1}
(\overline{\Omega})).$
\newline
{\bf B}) Let $1\le p\le 2$ and $ 1<\gamma<\frac{2p}{2-p}.$ Then
 $\partial_{\overline z}^{-1},\partial_{ z}^{-1}\in
\mathcal L(L^p( \Omega),L^\gamma(\Omega)).$
\newline
{\bf C})Let $1< p<\infty.$ Then  $\partial_{\overline z}^{-1},
\partial_{ z}^{-1}\in \mathcal L(L^p( \Omega),W^1_p(\Omega)).$
\end{proposition}
%
%
%
%
%
%
%
\section{Proof of Theorem 1}\label{sec2}

Let the function $\Phi$ satisfy (\ref{zzz}) and (\ref{mika}), and
$\widetilde x$ be some point from $\mathcal H\setminus\Gamma_0.$
Without loss of generality, adding to the function $\Phi$ a suitable
negative constant, we can always assume that
\begin{equation}\label{zion}
\max_{x\in\Omega}\varphi(x)<0.
\end{equation}

Let $a\in C^{6+\alpha}(\overline \Omega)$
be a holomorphic function such that
\begin{equation}\label{-5}
\frac{\partial a}{\partial \overline z}=0\quad \mbox{in}\,\,\Omega,
\quad \mbox{Re}\, a\vert_{\Gamma_0}=0.
\end{equation}

By Proposition 4.2 of \cite{IUY1}, there exists a holomorphic
function $a_0(z)\in C^7(\overline\Omega)$ such that $Im\,
a_0\vert_{\Gamma_0}=0,$ $a_0(\widetilde x)=1$ and $a_0$ vanishes at
each point of the set $\mathcal H\setminus \{\widetilde x\}$.
Then, choosing $\ell_0 \in \Bbb N$ large, we see that
$a=a_0^{\ell_0}$ is holomorphic with the following properties:

\begin{equation}\label{xoxo1} a(\widetilde x)= 1,\quad
\partial^{\alpha_1}_{x_1}\partial^{\alpha_2}_{x_2} a(x)
= 0 \quad \forall x\in \mathcal H\setminus\{\widetilde x\}\,\,\,
\mbox{and}\,\,\, \forall \alpha_1+\alpha_2\le 6.
\end{equation}
For example, we can choose $\ell_0=100$ and fix.
Short computations yield
\begin{equation}\label{oi}
L_1(x,D) (ae^{\tau\Phi})=q_1ae^{\tau\Phi},\quad L_1(x,D) (\overline
ae^{\tau\overline \Phi})=q_1\overline a e^{\tau\overline\Phi}.
\end{equation}

Let $e_1,e_2$ be smooth functions such that
\begin{equation}\label{short} e_1+e_2=1\quad
\mbox{on}\,\,\Omega,\quad
\end{equation}
and $e_1$ vanishes in a neighborhood of $\partial\Omega$ and $e_2$
vanishes in a neighborhood of the set $\mathcal H\setminus \overline
\Gamma_0.$

We have
\begin{proposition} \label{popo}
Let $q\in C^{2+\alpha}(\overline\Omega)$  for some positive $\alpha$
and $\widetilde q\in W^1_p(\overline\Omega)$ for some $p>2.$ Suppose
that $q\vert_{\mathcal H}=\widetilde q\vert_{\mathcal H}=0.$  There
exists smooth function $m_+\in C^2(\partial\Omega)$, which is
independent of $\tau$, such that the asymptotic formulae hold true:
\begin{equation}\label{50}
\widetilde{\mathcal R}_{\tau}(e_1(q+\frac{\widetilde
q}{\tau}))\vert_{\partial\Omega} = e^{\tau(\overline\Phi-\Phi)}\left
(\frac{m_+ e^{2i\tau\psi (\widetilde
x)}}{\tau^2}+o_{C^2(\partial\Omega)}(\frac{1}{\tau^2})\right)
\quad\mbox{as}\,\vert\tau\vert\rightarrow +\infty , \end{equation}
\begin{equation}\label{50s}
\quad{\mathcal R}_{\tau} (e_1(\overline q + \overline\frac{\widetilde
q}{\tau}))\vert_{\partial\Omega} =e^{\tau(\Phi-\overline\Phi)}\left
(\frac{\overline m_+ e^{-2i\tau\psi(\widetilde x)}}{\tau^2}
+o_{C^2(\partial\Omega)}(\frac{1}{\tau^2})\right
)\quad\mbox{as}\,\vert\tau\vert\rightarrow +\infty.
\end{equation}
and
\begin{equation}\label{robin}
\Vert \widetilde {\mathcal R}_{\tau}(e_1(q+\frac{\widetilde
q}{\tau}))-\frac{e_1q}{2\tau\partial_z\Phi}\Vert_{L^2(\Omega)}+\Vert
{\mathcal R}_{\tau}(e_1(\overline q + \overline\frac{\widetilde
q}{\tau}))-\frac{e_1\overline q}{2\tau\partial_{\overline
z}\overline\Phi}\Vert_{L^2(\Omega)}=o(\frac 1\tau)\quad
\mbox{as}\,\,\vert\tau\vert\rightarrow \infty.
\end{equation}
\end{proposition}

{\bf Proof.} Since $\mbox{supp}\, e_1\widetilde q\subset\subset \Omega$
by (\ref{short}),
the functions $\tau e^{\tau(\Phi-\overline \Phi)}\widetilde{\mathcal
R}_{\tau}(e_1\widetilde q)$ are uniformly bounded in
$C^k(\partial\Omega)$ for any positive integer $k.$
By Proposition 4 of \cite{IY1}, the functions $\tau e^{\tau(\Phi-\overline
\Phi)}\widetilde{\mathcal R}_{\tau}(e_1\widetilde q)$ converges to
zero pointwise on $\partial\Omega$ as $\tau$ approaches to
infinity. Therefore
\begin{equation}\label{5050}
\widetilde{\mathcal R}_{\tau}(e_1\frac{\widetilde
q}{\tau})\vert_{\partial\Omega} = e^{\tau(\overline\Phi-\Phi)}
o_{C^2(\partial\Omega)}(\frac{1}{\tau^2})
\quad\mbox{as}\,\vert\tau\vert\rightarrow +\infty .
\end{equation}
We set $q_*=\sum_{k=1}^\ell e(x-\widetilde x_\ell)
((\nabla q(\widetilde x_\ell),x-\widetilde x_\ell)
+ \frac 12(q''(\widetilde x_\ell)(x-\widetilde
x_\ell),(x-\widetilde x_\ell)))$, where $e$ is a smooth function
such that the support is located in a small ball centered at the origin
and $e$ is equal to one in some neighborhood of the origin.
Integrating by parts, we have
$$
\partial^{\beta_1}_{x_1}\partial^{\beta_2}_{x_2}\widetilde{\mathcal R}_{\tau}
(e_1(q-q_*))\vert_{\partial\Omega}
= \frac{1}{\pi\tau}\int_\Omega
\mbox{div}(\partial^{\beta_1}_{x_1}\partial^{\beta_2}_{x_2}
\frac{e_1(q-q_*)}{(\zeta-z)\vert\nabla\psi\vert^2}\nabla\psi)e^{2\tau
i\psi}d\xi\vert_{\partial\Omega},\quad \forall \beta_1+\beta_2\le 5.
$$
Since $
\mbox{div}(\partial^{\beta_1}_{x_1}\partial^{\beta_2}_{x_2}
\frac{e_1(q-q_*)}{(\zeta-z)\vert\nabla\psi\vert^2}\nabla\psi)$
belongs to the space $W^1_p(\Omega)$ with $p(\alpha)>2$, by
Proposition 4 of \cite{IY1} this integral converges to zero. Then
from the stationary phase argument, we obtain
$$
\widetilde{\mathcal R}_{\tau}(e_1q)\vert_{\partial\Omega} =
e^{\tau(\overline\Phi-\Phi)}\left (\frac{m_+ e^{2i\tau\psi
(\widetilde
x)}}{\tau^2}+o_{C^2(\partial\Omega)}(\frac{1}{\tau^2})\right)
\quad\mbox{as}\,\vert\tau\vert\rightarrow +\infty.
$$
Therefore the equality (\ref{50}) is proved.
The asymptotic formula (\ref{50s})
follows from (\ref{50}) and the equality $\quad{\mathcal R}_{\tau}
(e_1(\overline q + \overline\frac{\widetilde q}{\tau}))
= \overline{\widetilde{\mathcal R}_{\tau}(e_1
(q+\frac{\widetilde q}{\tau}))}.$
In order to prove (\ref{robin}), observe that by Proposition \ref{gandonnal},
the functions
$e^{2i\tau\psi}\widetilde {\mathcal R}_{\tau}(e_1\frac{\widetilde
q}{\tau})$ and $e^{-2i\tau\psi}{\mathcal
R}_{\tau}(e_1\overline\frac{\widetilde q}{\tau})$ converge
pointwise to zero and by Proposition \ref{Proposition 3.0} they
are bounded in $H^\alpha(\Omega)$ with some positive $\alpha$. Thus
they  converge to zero in $L^2(\Omega)$ as $\tau $ goes to infinity.
Applying Proposition 3.4 from \cite{IUY1}, we finish the proof
of (\ref{robin}). $\blacksquare$

Denote $p_1=\frac 12\partial^{-1}_{\overline z}(q_1a)-M(z)\in
C^{3+\alpha}(\overline \Omega)$, where the function $M$ is the polynomial
such that
\begin{equation}\label{kl}
p_1(\widetilde x)=0, \quad
\partial^{\alpha_1}_{x_1}\partial^{\alpha_2}_{x_2} p_1
(x)=0\quad \mbox{for $\forall x\in \mathcal H\setminus\{\widetilde x\}\,\,\,
\mbox{and}\,\,\, \forall \alpha_1+\alpha_2\le 3$}.
\end{equation}

Next we introduce holomorphic functions $a_{-1}, a_+
\in C^2(\overline \Omega)$ as follows:
\begin{equation}\label{zad-11}
(a_{-1}+\overline
a_{-1})\vert_{\Gamma_0}=\mbox{Re}\{\frac{p_1}{\partial_z\Phi}\},
\end{equation}
$$
\partial^{\alpha_1}_{x_1}\partial^{\alpha_2}_{x_2} a_{-1}(x)
=0\quad \mbox{for $\forall x\in \mathcal H\,\,\, \mbox{and}\,\,\, \forall
\alpha_1+\alpha_2\le 2$},
$$
\begin{equation}
(a_++\overline a_+)\vert_{\Gamma_0}=m_+ .
\end{equation}

We set $\widehat p_1=-q_1(\frac{e_1p_1}{2\partial_z\Phi}+a_{-1})+L_1(x,D)
(\frac{e_2p_1}{2\partial_z\Phi})$ and $\widetilde p_1=\frac
12\partial_{\overline z}^{-1}\hat p_1 -\widetilde M(z)$, where
$\widetilde M$ is a polynomial such that
\begin{equation}\label{gandon1}
\widetilde p_1(\widetilde x) =0,\quad
\partial^{\alpha_1}_{x_1}
\partial^{\alpha_2}_{x_2} \widetilde p_1
(x)=0\quad \mbox{for $\forall x\in \mathcal H\setminus\{\widetilde x\}\,\,\,
\mbox{and}\,\,\, \forall \alpha_1+\alpha_2\le 3$}.
\end{equation}
Since $\frac{\widetilde p_1}{2\partial_z\Phi}\in
H^1(\partial\Omega)$ by (\ref{short}) and (\ref{gandon1}), there exists
a holomorphic function $a_{-2} \in H^\frac 32(\overline\Omega)$ such
that
\begin{equation}\label{zad-111}
(a_{-2}+\overline a_{-2})\vert_{\Gamma_0}=\mbox{Re}\{\frac{\widetilde
p_1}{\partial_z\Phi}\}.
\end{equation}

By Proposition \ref{popo}, there exists a function $m_+\in
C^2(\partial\Omega)$ such that
\begin{equation}\label{50l}
\widetilde{\mathcal R}_{\tau}(e_1(p_1+\frac{\widetilde p_1}{\tau}))
= e^{\tau(\overline\Phi-\Phi)}\left (\frac{m_+ e^{2i\tau\psi
(\widetilde
x)}}{\tau^2}+o_{C^2(\partial\Omega)}(\frac{1}{\tau^2})\right)
\quad\mbox{as}\,\vert\tau\vert\rightarrow +\infty
\end{equation}
and
\begin{equation}\label{50ll}
\quad{\mathcal R}_{\tau} (e_1(\overline p_1 + \frac{\overline{\widetilde
p_1}}{\tau})) =e^{\tau(\Phi-\overline\Phi)} \left (\frac{\overline m_+
e^{-2i\tau\psi(\widetilde x)}}{\tau^2}
+o_{C^2(\partial\Omega)}(\frac{1}{\tau^2})\right
)\quad\mbox{as}\,\vert\tau\vert\rightarrow +\infty.
\end{equation}

We introduce the function $a_{\tau}\in H^1(\Omega)$ by
\begin{equation}\label{zad1}
a_{\tau}=a+\frac{a_{-1}-e_2p_1/2\partial_{
z}\Phi}{\tau}
+ \frac{1}{\tau^2}\left( e^{2i\tau\psi(\widetilde x)}a_+
+ e^{-2i\tau\psi(\widetilde x)}\overline a_++a_{-2}-\frac{\widetilde
p_1 e_2}{2\partial_z\Phi}\right).
\end{equation}

Using this formula, we prove the following proposition.
\begin{proposition}\label{Proposition 00}
The asymptotic formulae are true:
\begin{eqnarray}\label{249}
L_1(x,D)(a_{\tau}e^{\tau \Phi}+\overline a_{\tau} e^{\tau \overline
\Phi}-e^{\tau\Phi}\widetilde{\mathcal R}_{\tau} (e_1(p_1+\widetilde
p_1/\tau))-e^{\tau\overline\Phi}{\mathcal R}_{\tau}(e_1(\overline
p_1+\overline{\widetilde
p_1}/\tau)))=e^{\tau\varphi}o_{L^2(\Omega)}(\frac{1}{\tau}) ,\\
\quad (a_{\tau}e^{\tau \Phi}+\overline a_{\tau} e^{\tau \overline
\Phi}-e^{\tau\Phi}\widetilde{\mathcal R}_{\tau}(e_1(p_1+\widetilde
p_1/\tau))-e^{\tau\overline\Phi}{\mathcal R}_{\tau}(e_1(\overline
p_1+\overline{\widetilde
p_1}/\tau)))\vert_{\Gamma_0}=e^{\tau\varphi}o_{H^1(\Gamma_0)} (\frac
{1}{\tau^2}).
\end{eqnarray}
\end{proposition}

{\bf Proof.} By (\ref{zzz}) and (\ref{zad-11})-(\ref{zad-111}), we have
$$
(a_{\tau}e^{\tau \Phi}+\overline a_{\tau} e^{\tau \overline
\Phi}-e^{\tau\Phi}\widetilde{\mathcal R}_{\tau}(e_1(q_1+\widetilde
p_1/\tau))-e^{\tau\overline\Phi}{\mathcal R}_{\tau }(e_1(\overline
p_1+\overline{\widetilde p_1}/\tau))\vert_{\Gamma_0}
$$
$$
= (a_{\tau}e^{\tau \varphi}+\overline a _{\tau} e^{\tau
\varphi}-e^{\tau\varphi}\widetilde{\mathcal R}_{\tau}(e_1(p_1+\overline
p_1/\tau))-e^{\tau\varphi}{\mathcal R}_{\tau}(e_1(\overline
p_1+\overline{\widetilde p_1}/\tau))\vert_{\Gamma_0}
$$
$$
= e^{\tau\varphi}(a+\frac{a_{-1}-e_2p_1/2\partial_{ z}\Phi}{\tau}
+\frac{1}{\tau^2}( e^{2i\tau\psi(\widetilde
x)}a_++e^{-2i\tau\psi(\widetilde x)}\overline a_+
+a_{-2}-\frac{\widetilde p_1e_2}{2\partial_z\Phi}) $$
$$
+ \overline a+\frac{\overline a_{-1}-e_2 \overline p_1/2\partial_{\overline
z}\overline\Phi}{\tau}+\frac{1}{\tau^2}( e^{2i\tau\psi(\widetilde
x)}a_+ +e^{-2i\tau\psi(\widetilde x)}\overline a_++\overline a_{-2}
-\frac{\overline{\widetilde p_1}e_2}{2\partial_{\overline
z}\overline\Phi})
$$
$$
-\widetilde{\mathcal R}_{\tau}(e_1(p_1+\widetilde
p_1/\tau))-{\mathcal R}_{\tau}(e_1(\overline p_1+\overline{\widetilde
p_1}/\tau)))\vert_{\Gamma_0}
$$
$$
= e^{\tau\varphi}\{\frac{1}{\tau^2} (e^{2i\tau\psi(\widetilde
x)}a_++e^{-2i\tau\psi(\widetilde x)}\overline a_+ +
e^{2i\tau\psi(\widetilde x)}a_++e^{-2i\tau\psi(\widetilde x)}\overline
a_+)
$$
$$
- \widetilde{\mathcal R}_{\tau}(e_1(p_1+\widetilde
p_1/\tau))-{\mathcal R}_{\tau}(e_1(\overline
p_1+\overline{\widetilde
p_1}/\tau))\}\vert_{\Gamma_0}=e^{\tau\varphi}o_{H^1(\Gamma_0)}(\frac
{1}{\tau^2}).
$$
Here in order to obtain the final equality, we used (\ref{50l}) and
(\ref{50ll}).  Proposition \ref{popo} and simple computations
imply the asymptotic formula:
\begin{eqnarray}\label{251}
L_1(x,D)(-e^{\tau\Phi}\widetilde{\mathcal
R}_{\tau}(e_1(p_1+\widetilde p_1/\tau))-\frac{e_2(p_1+\widetilde
p_1/\tau)e^{\tau\Phi}}{2\tau\partial_z\Phi}-e^{\tau\overline\Phi}{\mathcal
R}_{\tau}(e_1(\overline p_1+\overline{\widetilde
p_1}/\tau))\nonumber\\-\frac{e_2(\overline p_1+\overline{\widetilde
p_1}/\tau)e^{\tau\overline\Phi}}{2\tau\partial_{\overline
z}\overline\Phi})
                                                      \nonumber
= -L_1(x,D)(e^{\tau\Phi}\widetilde{\mathcal R}_{\tau}
(e_1(p_1+\widetilde p_1/\tau))+\frac{e_2(p_1+\widetilde
p_1/\tau)e^{\tau\Phi}}{2\tau\partial_z\Phi})\\
- L_1(x,D)(e^{\tau\overline\Phi}{\mathcal R}_{\tau}(e_1(\overline
p_1+\overline{\widetilde p_1}/\tau))+\frac{e_2(\overline
p_1+\overline{\widetilde p_1}/\tau)
e^{\tau\overline\Phi}}{2\tau\partial_{\overline z}\overline\Phi})\nonumber\\
= -q_1e^{\tau\Phi}\widetilde {\mathcal R}_{\tau}(e_1(p_1+\widetilde
p_1/\tau))-q_1e^{\tau\overline\Phi}{\mathcal R}_{\tau}(e_1(\overline
p_1+\overline{\widetilde
p_1}/\tau))\nonumber\\
-e^{\tau\Phi}L_1(x,D) (\frac{e_2(p_1+\widetilde
p_1/\tau)}{2\tau\partial_z\Phi})-e^{\tau\overline\Phi}L_1(x,D)
(\frac{e_2(\overline p_1+\overline{\widetilde
p_1}/\tau)}{2\tau\partial_{\overline z}\overline\Phi})
\nonumber\\-q_1\overline a e^{\tau\overline\Phi}-q_1 a e^{\tau\Phi}
-q_1e^{\tau\Phi}\frac{e_1p_1} {2\tau\partial_z\Phi}-q_1
e^{\tau\overline\Phi}\frac{e_1\overline p_1}{2\tau\partial_{\overline z}
\overline\Phi}\nonumber\\
+ \frac 1\tau(q_1\frac{e_1p_1}{2\partial_z\Phi}+L_1(x,D)
(\frac{e_2p_1}{2\partial_z\Phi}))e^{\tau\Phi} + \frac
1\tau(q_1\frac{e_1\overline p_1}{2\partial_{\overline
z}\overline\Phi}+L_1(x,D) (\frac{e_2\overline p_1}{2\partial_{\overline
z}\overline\Phi}))e^{\tau\overline\Phi}+e^{\tau\varphi}o_{L^2(\Omega)}(\frac{1}{\tau})\nonumber\\
= - \frac{1}{\tau}q_1 a_{-1}e^{\tau\Phi}-\frac{1}{\tau}q_2\overline
a_{-1}e^{\tau\overline\Phi} - q_1\overline a e^{\tau\overline\Phi}-q_2 a
e^{\tau\Phi} +e^{\tau\varphi}o_{L^2(\Omega)}(\frac{1}{\tau}).
\end{eqnarray}
Similarly to (\ref{oi}), we obtain
\begin{eqnarray}\label{250}
L_1(x,D)(a_{\tau}e^{\tau\Phi}+\overline
a_{\tau}e^{\tau\overline\Phi}-\frac{e_2(p_1+\widetilde
p_1/\tau)e^{\tau\Phi}}{2\tau\partial_z\Phi}-\frac{e_2(\overline
p_1+\overline{\widetilde
p_1}/\tau)e^{\tau\overline\Phi}}{2\tau\partial_{\overline
z}\overline\Phi})\nonumber\\
= q_1 (a_{\tau}-\frac{e_2(p_1+\widetilde
p_1/\tau)}{2\tau\partial_z\Phi})e^{\tau\Phi}
+ q_1(\overline a_{\tau} -\frac{e_2(\overline p_1+\overline{\widetilde
p_1}/\tau)}{2\tau\partial_{\overline
z}\overline\Phi})e^{\tau\overline\Phi}.
\end{eqnarray}
By (\ref{250}) and (\ref{251}), we obtain (\ref{249}). $\blacksquare$

Using Propositions \ref{vanka} and \ref{Proposition 00},
we construct the last term $u_{-1}$ in the complex geometric
optics solution which satisfies
\begin{equation}\label{mimino11}
\Vert u_{-1}\Vert_{H^{2,\tau}(\Omega)}/\vert\tau\vert^\frac
32=o(\frac{1}{\tau})\quad \mbox{as}\,\,\vert\tau\vert\rightarrow
+\infty.
\end{equation}
Finally we obtain a complex geometric optics solution
for the linear operator $L_1(x,D)$ in the form:
\begin{equation}\label{zad}
u_{1,*}(x)=a_{\tau}e^{\tau \Phi}+\overline a_{\tau} e^{\tau \overline
\Phi}-e^{\tau\Phi}\widetilde{\mathcal R}_{\tau}(p_1+\widetilde
p_1/\tau)-e^{\tau\overline\Phi}{\mathcal R}_{\tau}(\overline
p_1+\overline{\widetilde p_1}/\tau)+e^{\tau \varphi} u_{-1}.
\end{equation}

Obviously
\begin{equation}\label{zad22}
L_1(x,D)u_{1,*}=0\quad\mbox{in}\,\,\Omega,\quad
u_{1,*}\vert_{\Gamma_0}=0.
\end{equation}

Thanks to (\ref{01}), there exist positive constants $C$ and
$\kappa$, both independent of $\tau,$ such that
\begin{equation}\label{verevka}
\Vert e^{-\tau\varphi}P_1(x,D)u_{1,*}\Vert_{L^2(\Omega)}\le
Ce^{-\kappa\tau}.
\end{equation}

We finish the construction of the complex geometric optics solution
for the semilinear elliptic equation $P_1(x,D)=L_1(x,D)-f_1$ using
the Newton-Kantorovich iteration scheme. More precisely we use the
Theorem 6 (1.XVIII) from \cite{AK} p.708.

Following the notations of \cite{AK}, we set $x_0=0, X=\{u\in
H^{2,\tau}(\Omega),u\vert_{\Gamma_0}=0\}, \Vert\cdot\Vert_X
=\Vert\cdot\Vert=\Vert\cdot\Vert_{H^{2,\tau}(\Omega)}$
and $P=I+L_1(x,D+i\tau\nabla
\varphi)^{-1}e^{-\tau\varphi}f(x, e^{\tau\varphi}\circ)$. Here by
$L_1(x,D+i\tau\nabla \varphi)^{-1}$ we mean the operator from
$L^2(\Omega) $ into the orthogonal complement of
$\mbox{Ker}\,L_1(x,D+i\tau\nabla \varphi)$ in $ X,$ and $I$ is the
identity operator. The mapping $P$ is twice continuously
differentiable as the mapping from $X$ into $X.$ By Proposition
\ref{vanka}, we have
$$ \Vert \Gamma_0\Vert_{\mathcal
L(X;X)}\le C\tau^2. $$

 From this
inequality and (\ref{verevka}), we have
$$
\Vert \Gamma_0 P(x_0)\Vert\le C\tau^2e^{-\kappa\tau}=\eta(\tau).
$$
We set $\Omega_0=\{x\vert\Vert x-x_0\Vert\le r_0\}.$ By (\ref{02}),
we have
$$
\Vert \Gamma_0P''(x)\Vert\le C\tau^2=K(\tau).
$$
Then $h=K(\tau)\eta\le \tau^4e^{-\kappa\tau}$ and
$r_0=\frac{1-\root\of{1-2h}}{h}\eta\le 2\tau^2e^{-\kappa\tau}<\frac
12 $ for all sufficiently large $\tau.$ Then there exists a solution
$x_*$ to the equation  $P(x)=0$ such that $\Vert x_*\Vert\le r_0.$

Let $u_{1}$ be a complex geometrical optics solution to the
semilinear equation $P_1(x,D)$ of the form:
\begin{equation}\label{pobeda}
u=u_{1,*}+e^{\tau \varphi}u_{cor},\quad \Vert u_{cor}\Vert
_{H^{2,\tau}(\Omega)}=o(\frac 1\tau)\quad
\mbox{as}\,\,\tau\rightarrow +\infty.
\end{equation}

Similarly we construct the complex geometric optics solutions to the
operator $L_2(x,D):$
\begin{equation}\label{zads}
v(x)=\widetilde a_{\tau}e^{-\tau \Phi}+\overline{\widetilde a}_{\tau}
e^{-\tau \overline \Phi}-e^{-\tau\Phi}\widetilde{\mathcal
R}_{-\tau}(p_2+\widetilde p_2/\tau)-e^{-\tau\overline\Phi}{\mathcal
R}_{-\tau}(\overline p_2+\overline{\widetilde p_2}/\tau)+e^{-\tau
\varphi} v_{-1}.
\end{equation}

Here the function $\widetilde a_{\tau}$ is given by
\begin{equation}\label{zad1}
\widetilde a_{\tau}=a+\frac{\widetilde a_{-1}+e_2p_2/2\partial_{
z}\Phi}{\tau}+\frac{1}{\tau^2}\left( e^{2i\tau\psi(\widetilde
x)}a_-+e^{-2i\tau\psi(\widetilde x)}\overline a_-+\widetilde
a_{-2}-\frac{\widetilde p_2 e_2}{2\partial_z\Phi}\right),
\end{equation}

where $ p_2=\frac 12\partial^{-1}_{\overline z}(q_2a)-M_1(z)\in
C^{3+\alpha}(\overline \Omega)$ and the function $M_1$ is the polynomial
such that
\begin{equation}\label{kl}
p_2(\widetilde x)=0, \quad
\partial^{\alpha_1}_{x_1}\partial^{\alpha_2}_{x_2} p_2
(x)=0\quad \mbox{for $\forall x\in \mathcal H\setminus\{\widetilde x\}\,\,\,
\mbox{and}\,\,\, \forall \alpha_1+\alpha_2\le 3$.}
\end{equation}

The function $\widetilde a_{-1}\in C^2(\overline \Omega)$ is the
holomorphic functions such that :
\begin{equation}\label{zad-1}
(\widetilde a_{-1}+\overline{\widetilde
a_{-1}})\vert_{\Gamma_0}=\mbox{Re}\{\frac{p_2}{\partial_z\Phi}\},
\end{equation}
$$
\partial^{\alpha_1}_{x_1}\partial^{\alpha_2}_{x_2} \widetilde a_{-1}(x)
=0\quad \mbox{for $\forall x\in \mathcal H\,\,\, \mbox{and}
\,\,\, \forall \alpha_1+\alpha_2\le 2$}.
$$

We set  $\widetilde p_2=\frac 12\partial_{\overline z}^{-1}\hat p_2 -\widetilde
M_1(z)$ and $\widehat p_2=-q_2(\frac{e_1p_2}{2\partial_z\Phi}+\widetilde
a_{-1})+L_2(x,D) (\frac{e_2p_2}{2\partial_z\Phi})$, where $\widetilde
M_1$ is a polynomial such that
\begin{equation}\label{gandon11}
\widetilde p_2(\widetilde x) =0,\quad
\partial^{\alpha_1}_{x_1}
\partial^{\alpha_2}_{x_2} \widetilde p_2
(x)=0\quad \mbox{for $\forall x\in \mathcal H\setminus\{\widetilde x\}\,\,\,
\mbox{and}\,\,\, \forall \alpha_1+\alpha_2\le 2$}.
\end{equation}
Since $\frac{\widetilde p_2}{2\partial_z\Phi}\in
H^1(\partial\Omega)$ by (\ref{gandon11}), there exists a holomorphic
function $\widetilde a_{-2} \in H^1(\overline\Omega)$ such that
\begin{equation}\label{zad-1}
(\widetilde a_{-2}+\overline{\widetilde
a_{-2}})\vert_{\Gamma_0}=\mbox{Re}\{\frac{\widetilde
p_2}{\partial_z\Phi}\}.
\end{equation}
By Proposition \ref{popo}, there exists a function $m_-\in
C^2(\partial\Omega)$ such that
\begin{equation}\label{50l}
\widetilde{\mathcal R}_{\tau}(e_1(p_2+\frac{\widetilde p_2}{\tau}))
= e^{\tau(\overline\Phi-\Phi)}\left (\frac{m_- e^{2i\tau\psi
(\widetilde
x)}}{\tau^2}+o_{C^2(\partial\Omega)}(\frac{1}{\tau^2})\right)
\quad\mbox{as}\,\vert\tau\vert\rightarrow +\infty
\end{equation}
and
\begin{equation}\label{50ll}
\quad{\mathcal R}_{\tau} (e_1(\overline p_2 + \frac{\overline{\widetilde
p_2}}{\tau})) =e^{\tau(\Phi-\overline\Phi)} \left (\frac{\overline m_-
e^{-2i\tau\psi(\widetilde x)}}{\tau^2}
+o_{C^2(\partial\Omega)}(\frac{1}{\tau^2})\right
)\quad\mbox{as}\,\vert\tau\vert\rightarrow +\infty.
\end{equation}

Next we introduce a holomorphic function $ a_-\in C^2(\overline
\Omega)$ such that :

\begin{equation}
(a_-+\overline a_-)\vert_{\Gamma_0}=m_- .
\end{equation}

Obviously the function $\widetilde a_{\tau}$ belongs to $H^1(\Omega).$
Using Proposition \ref{popo}, we have
\begin{eqnarray}\label{90}
{L}_2(x,{D})\left(-e^{-\tau\Phi}\widetilde{\mathcal R}
_{-\tau}(e_1(p_2+\frac{\widetilde
p_2}{\tau}))+\frac{e^{-\tau\Phi}e_2(p_2+\frac{\widetilde
p_2}{\tau})}{2\tau\partial_z\Phi}\right.\nonumber\\
\left.-e^{-\tau\overline\Phi}{\mathcal R}_{-\tau } (e_1(\overline
p_2+\frac{\overline{\widetilde
p_2}}{\tau}))+\frac{e^{-\tau\overline\Phi}e_2(\overline
p_2+\frac{\overline{\widetilde
p_2}}{\tau})}{2\tau\partial_{\overline z}\overline\Phi}\right )
\nonumber\\
= -{L}_2(x,{D})\left (e^{-\tau\Phi}\widetilde{\mathcal
R}_{-\tau}(e_1(p_2+\frac{\widetilde
p_2}{\tau}))-\frac{e^{-\tau\Phi}(e_2(p_2+\frac{\widetilde
p_2}{\tau})}{2\tau\partial_z\Phi}\right)                     \nonumber\\
- {L}_2(x,{D})\left (e^{-\tau\overline\Phi}{\mathcal
R}_{-\tau}(e_1(\overline p_2+\frac{\overline{\widetilde
p_2}}{\tau}))-\frac{e^{-\tau\overline\Phi}e_2(\overline
p_2+\frac{\overline{\widetilde
p_2}}{\tau})}{2\tau\partial_{\overline z}\overline\Phi}\right )
                             \nonumber\\
= -e^{-\tau\Phi}q_2\widetilde{\mathcal
R}_{-\tau}(e_1(p_2+\frac{\widetilde
p_2}{\tau}))
-e^{-\tau\overline\Phi}q_2{\mathcal R}_{-\tau}(e_1(\overline
p_2+\frac{\overline{\widetilde p_2}}{\tau}))
\nonumber\\
-q_2(a+\frac{\widetilde a_{-1}}{\tau})e^{-\tau\Phi} -q_2(\overline
a+\frac{\overline{\widetilde a_{-1} }}{\tau}) e^{-\tau\overline\Phi}+
e^{-\tau\varphi}o_{L^2(\Omega)}(\frac{1}{\tau})\quad \mbox{as}
\,\,\tau\rightarrow +\infty.
\end{eqnarray}

Setting $v^*=\widetilde a_\tau e^{-\tau \Phi}+\overline{\widetilde a_\tau}
e^{-\tau \overline \Phi}-e^{-\tau\Phi}\widetilde{\mathcal
R}_{-\tau}(e_1(p_2+\frac{\widetilde
p_2}{\tau}))-e^{-\tau\overline\Phi}{\mathcal R}_{-\tau}(e_1(\overline
p_2+\frac{\overline{\widetilde p_2}}{\tau}))$, we obtain that
\begin{equation}\label{nino}
L_2(x,D)v^*=e^{-\tau\varphi}o_{L^2(\Omega)}(\frac
1\tau)\quad\mbox{in}\,\,\Omega, \quad
v^*\vert_{\Gamma_0}=e^{-\tau\varphi}o_{H^1(\Gamma_0)}(\frac
1\tau)\quad\mbox{as}\,\,\tau\rightarrow +\infty.
\end{equation}

Using (\ref{nino}) and Proposition \ref{vanka} and \ref {Proposition
00}, we construct the last term $v_{-1}\in H^2(\overline \Omega)$ in the
complex geometric optics solution which solves the boundary value
problem
\begin{equation}\label{nino1}
L_2(x,D)v_{-1}=L_2(x,D)v^* \quad\mbox{in}\,\,\Omega, \quad
v_{-1}\vert_{\Gamma_0}=v^*,
\end{equation}
and we obtain
\begin{equation}\label{Amimino11}
\root\of{\vert\tau\vert} \Vert v_{-1} \Vert_{L^2(\Omega)} +
\frac{1}{\root\of{\vert\tau\vert} } \Vert (\nabla v_{-1})
\Vert_{L^2(\Omega)}=o(\frac{1}{\tau})\quad\mbox{as}
\,\,\tau\rightarrow +\infty.
\end{equation}
Finally we have a complex geometric optics solution for
the Schr\"odinger operator $L_2(x,D)$ in a form:
\begin{equation}\label{-3}
v=v^*+v_{-1}e^{-\tau \varphi}.
\end{equation}


By (\ref{-3}), (\ref{nino}) and (\ref{nino1}), we have
\begin{equation}\label{-4}
L_2(x,D)v=0\quad\mbox{in}\,\,\Omega, \quad v\vert_{\Gamma_0}=0.
\end{equation}

Let $u_2$ be a solution to the following boundary value
problem:
\begin{equation}\label{(2.1I)}
{ L}_{2}(x,D)u_2-f(x,u_2)=0\quad \mbox{in}\,\,\Omega,\quad
u_2\vert_{\partial\Omega}=u_1\vert_{\partial\Omega}, \quad
\frac{\partial u_2}{\partial \nu}\vert_{\widetilde \Gamma}
=\frac{\partial u_1}{\partial\nu}\vert_{\widetilde \Gamma}.
\end{equation}
Taking the scalar products of equation (\ref{(2.1I)}) with the function
$u_2$ and integrating by parts, we have
\begin{equation}\label{(2.1II)}
\int_\Omega (\vert \nabla u_2\vert^2+C_2\vert u_2\vert^{p+1})dx\le
\int_{\widetilde\Gamma} u_2\frac{\partial
u_2}{\partial\nu}d\sigma+\int_\Omega q_2 u_2^2dx
+C\,\mbox{Vol}(\Omega).
\end{equation}

From (\ref{(2.1II)}), using (\ref{2}), we have
\begin{equation} \label{xxx}
\Vert u_2\Vert_{H^1(\Omega)}\le C.
\end{equation}
Then by (\ref{00}) and (\ref{02}), there exists $q_3\in L^p(\Omega)$  for any $p\in
(1,\infty)$ such that
$$
(\Delta +q_3)u_2=0\quad\mbox{in}\,\,\Omega, \quad
u_2\vert_{\partial\Omega}=u_1\vert_{\partial\Omega}, \quad
\frac{\partial u_2}{\partial \nu}\vert_{\widetilde \Gamma}
=\frac{\partial u_1}{\partial\nu}\vert_{\widetilde \Gamma}.
$$

Applying to this equation Carleman estimate (\ref{suno4}), we obtain
\begin{equation}\label{xxx1}
\Vert u_2e^{-\tau\varphi}\Vert_{H^{1,\tau}(\Omega)}\le C\vert
\tau\vert^\frac 12\quad\forall \tau\ge \tau_0.
\end{equation}

Similarly
\begin{equation}\label{xxx11}
\Vert u_1e^{-\tau\varphi}\Vert_{H^{1,\tau}(\Omega)}\le C\vert
\tau\vert^\frac 12\quad\forall \tau\ge \tau_0.
\end{equation}
By (\ref{zad1}), (\ref{mimino11}) and (\ref{pobeda}), we have
\begin{equation}\label{xxx111}
\Vert u_1 e^{-\tau\varphi}\Vert_{H^\frac 32(\partial \Omega)}\le C
\tau^2.
\end{equation}
Hence, by (\ref{xxx1}), (\ref{xxx11}) and (\ref{xxx111}) we obtain
\begin{equation}\label{zion1}
\Vert u_1e^{-\tau\varphi}\Vert_{H^{1,\tau}(\Omega)}+\Vert
u_2e^{-\tau\varphi}\Vert_{H^{1,\tau}(\Omega)}\le C\vert
\tau\vert^2\quad\forall \tau\ge \tau_0.
\end{equation}
Therefore, by (\ref{zion1}) and (\ref{zion}), there exists
$\tau_1>0$ such that
\begin{equation}\label{ZZ}
\Vert u_1\Vert_{C^0(\overline \Omega)}+\Vert u_2\Vert_{C^0(\overline
\Omega)}\le \delta\quad \forall \tau\ge \tau_1.
\end{equation}

Setting $u=u_1-u_2$, we have
\begin{equation}
{L}_2(x,{D})u
+(q_1-q_{2})u_1 +f_1(x,u_1)-f_2(x,u_2)=0 \quad \mbox{in}~ \Omega    \label{mn}
\end{equation}
and
\begin{equation}\label{mn1}
u \vert_{\partial\Omega} =0, \quad \frac{\partial u}{\partial \nu}
\vert_{\widetilde \Gamma} =0.
\end{equation}

Let $v$ be a function given by  (\ref{-3}).  Taking the scalar
products of (\ref{mn}) with $ v$  in $L^2(\Omega)$ and using
(\ref{-4}) and (\ref{mn1}), we obtain
\begin{equation}\label{ippolit}
0=\frak G(u_1,v)= \int_{\Omega}(q_1-q_{2})u_1  v dx+\int_\Omega
(f_1(x,u_1)-f_2(x,u_2))v dx.
\end{equation}
Our goal is to obtain the asymptotic formula for the right-hand side
of (\ref{ippolit}).

By (\ref{00}) and (\ref{01}), there exist positive constants $C$
and $\delta$ such that
\begin{equation}\label{beluga}
\vert f(x,y)\vert\le C\vert y\vert^p,\quad \forall
(x,y)\in\Omega\times [-\delta,\delta].
\end{equation}

Using (\ref{beluga}), (\ref{ZZ}) and (\ref{zion}), we obtain
\begin{eqnarray}
\vert \int_\Omega (f_1(x,u_1)-f_2(x,u_2))v dx\vert
\le \int_\Omega (\vert f_1(x,u_1)\vert+\vert f_2(x,u_2)\vert)
\vert v\vert dx \le C\int_\Omega (\vert u_1\vert^{p}
+\vert u_2\vert^{p})\vert v\vert dx \nonumber\\
\le C\int_\Omega e^{(p-1)\tau\varphi}
(\vert e^{-\tau\varphi}u_1\vert^{p}+\vert e^{-\tau\varphi}
u_2\vert^{p})\vert e^{\tau\varphi} v\vert dx\le
 Ce^{p\tau \max_{x\in\Omega}\varphi}=o(\frac{1}{\tau}).
\end{eqnarray}
By (\ref{pobeda}), (\ref{zad}),  (\ref{mimino11}) and Proposition
\ref{popo}, we have
\begin{equation}\label{P1}
u_1(x)=2\mbox{Re}\,\{(a+\frac{a_{-1}}{\tau})e^{\tau\Phi}-\frac{p_1e^{\tau\Phi}}
{2\tau{\partial_z\Phi}}\}+e^{\tau\varphi}o_{L^2(\Omega)} (\frac
1\tau)\,\,\mbox{as}\,\,\tau\rightarrow +\infty.
\end{equation}
Using (\ref{zads}), (\ref{Amimino11}) and Proposition \ref{popo},
we obtain
\begin{equation}\label{P2}
v(x)=2\mbox{Re}\,\{(a+\frac{\widetilde
a_{-1}}{\tau})e^{-\tau\Phi}+\frac{p_2e^{-\tau\Phi}}
{2\tau{\partial_z\Phi}}\}+e^{-\tau\varphi}o_{L^2(\Omega)}(\frac
1\tau)\,\,\mbox{as}\,\,\tau\rightarrow +\infty.
\end{equation}

By (\ref{P1}) and (\ref{P2}), we obtain the following asymptotic formula:
\begin{eqnarray}\label{nonsence}
\frak G(u_1,v)=((q_1-q_{2})u_1,v)_{L^2(\Omega)} =
((q_1-q_{2})((a+\frac{a_{-1}}{\tau})e^{\tau\Phi}+({\overline
a}+\frac{\overline a_{-1}}{\tau})e^{\tau\overline\Phi}-\frac{\overline
p_1e^{\tau\overline\Phi}}
{2\tau\overline{\partial_z\Phi}}-\frac{p_1e^{\tau\Phi}}
{2\tau{\partial_z\Phi}}+e^{\tau\varphi}o_{L^2(\Omega)}(\frac
1\tau)),
\nonumber\\
(a+\frac{\widetilde a_{-1}}{\tau})e^{-\tau\Phi}+({\overline
a}+\frac{\overline{\widetilde
a_{-1}}}{\tau})e^{-\tau\overline\Phi}+\frac{\overline p_2
e^{-\tau\overline\Phi}} {2\tau\overline{\partial_z\Phi}}+\frac{p_2
e^{-\tau\Phi}}
{2\tau{\partial_z\Phi}}+e^{-\tau\varphi}o_{L^2(\Omega)} (\frac
1\tau))
_{L^2(\Omega)}                              \nonumber\\
= \int_\Omega  (2(q_1-q_{2})
\mbox{Re}\{(a+\frac{a_{-1}}{\tau}-\frac{p_1}
{2\tau{\partial_z\Phi}})(a+\frac{\widetilde a_{-1}}{\tau}+\frac{p_2}
{2\tau{\partial_z\Phi}})\}+2(q_1-q_2)\mbox{Re} \{\vert a\vert^2
e^{2\tau i\psi}\})dx+o(\frac 1\tau).
                                 \nonumber
\end{eqnarray}
Applying the stationary phase argument (see e.g., \cite{BH}) to the
last integral on the right-hand side of this formula and using
(\ref{xoxo1}), we have
\begin{eqnarray}\label{sosika1}\frak G(u_1,v)=\int_\Omega 2(q_1-q_{2})
\mbox{Re}\{(a+\frac{a_{-1}}{\tau}-\frac{p_1}
{2\tau{\partial_z\Phi}})(a+\frac{\widetilde a_{-1}}{\tau}+\frac{p_2}
{2\tau{\partial_z\Phi}})\}dx\\ +2\pi \frac{(q_1-q_{2})(\widetilde x)
e^{2\tau i\psi(\widetilde x)}
+(q_1-q_{2})(\widetilde x)e^{-2i\tau\psi(\widetilde
x)}} {\tau\vert \mbox{det}\, \psi ''(\widetilde x)\vert^\frac 12}
                                               \nonumber\\
+\frac {1}{2\tau i}\int_{\partial\Omega} (q_1-q_{2})\vert
a\vert^2 e^{2\tau i\psi}
\frac{(\nu,\nabla\psi)}{\vert\nabla\psi\vert^2}d\sigma - \frac
{1}{2\tau i}\int_{\partial\Omega} (q_1-q_{2})\vert a\vert^2
e^{-2\tau i\psi}
\frac{(\nu,\nabla\psi)}{\vert\nabla\psi\vert^2}d\sigma+o(\frac
1\tau).\nonumber
\end{eqnarray}
By Proposition \ref{Proposition -1}, we have
\begin{equation}\label{sosiska}
\frac {1}{2\tau i}\int_{\partial\Omega} (q_1-q_{2})\vert
a\vert^2 e^{2\tau i\psi}
\frac{(\nu,\nabla\psi)}{\vert\nabla\psi\vert^2}d\sigma - \frac
{1}{2\tau i}\int_{\partial\Omega} (q_1-q_{2})\vert a\vert^2
e^{-2\tau i\psi}
\frac{(\nu,\nabla\psi)}{\vert\nabla\psi\vert^2}d\sigma=
o(\frac{1}{\tau}).
\end{equation}
Since $\psi(\widetilde x)\ne 0$, we obtain from (\ref{sosika1})
and (\ref{sosiska}) that $q_1(\widetilde x)=q_2(\widetilde x).$
Since $\widetilde x$ can be chosen an arbitrary close to any point
in the domain $\Omega$, we finish the proof.$\blacksquare$
\\
\vspace{0.3cm}
\\

{\bf Acknowledgements.}
Most part of the paper has been written during the stay of
the first named author at Graduate School of Mathematical
Sciences of The University of Tokyo and he thanks the Global
COE Program ``The Research and Training Center for New Development
in Mathematics" for support of the visit to The University of Tokyo.

\end{document}